\input{aipcheck}

\documentclass[
    ,final            
  ]
  {aipproc}

\layoutstyle{6x9}

\begin{document}

\title{GRB\,980425 in the Off-Axis Jet Model of the Standard GRBs}

\author{Ryo Yamazaki}{
  address={Department of Physics, 
           Kyoto University, Kyoto 606-8502, Japan}
}

\author{Daisuke Yonetoku}{
  address={Department of Physics, Kanazawa University, 
           Kakuma, Kanazawa, Ishikawa 920-1192, Japan}
}

\author{Takashi Nakamura}{
  address={Department of Physics, 
           Kyoto University, Kyoto 606-8502, Japan}
}

\def\E{{E}}
\def\lesssim{{\leq}}
\def\gtrsim{{\geq}}
\def\degr{{^\circ}}
\def\N{\nonumber}
\def\f{\frac}

\begin{abstract}
Using a simple off-axis jet model of GRBs,
we can  reproduce the observed unusual properties of the prompt
emission of GRB\,980425, such as the extremely low isotropic 
equivalent $\gamma$-ray energy, the low peak energy, 
the high fluence ratio, and the long  spectral lag when 
the jet with the standard energy of $\sim10^{51}$~ergs
and the opening half-angle of 
$10\degr\lesssim\Delta\theta\lesssim30\degr$
is seen from the off-axis viewing angle
$\theta_v\sim \Delta \theta +10\gamma^{-1}$,
where $\gamma$ is a Lorentz factor of the jet.
For our adopted fiducial parameters,
if the jet that caused GRB\,980425 is viewed from 
the on-axis direction,
the intrinsic peak energy $E_p(1+z)$ is $\sim$2.0--4.0~MeV,
which corresponds to those of GRB\,990123 and GRB\,021004.
Our model might be able to explain the other unusual properties
of this event.
We also discuss the connection of GRB\,980425 in our model 
with the X-ray flash, and
the origin of a class of GRBs with small $E_\gamma$
such as GRB\,030329.

\end{abstract}

\maketitle

\section{INTRODUCTION}
There are some GRBs that were thought to be 
associated with SNe \citep{dv03,stanek}.
GRB\,980425 / SN\,1998bw, located at $z=0.0085$ (36~Mpc), was the 
first event of such class \citep{ga98,kul98,pian00,pian03}.
%
%
It is important to investigate whether 
GRB\,980425 is similar to more or less 
typical long duration  GRBs.
However, GRB\,980425 showed unusual observational properties.
The isotropic equivalent $\gamma$-ray energy
is $E_{iso}\sim 6\times10^{47}$~ergs and
the geometrically corrected energy is 
$E_\gamma=(\Delta\theta)^2E_{iso}/2\sim3
\times10^{46}$\,ergs\,$(\Delta\theta/0.3)^2$, where
$\Delta\theta$ is the unknown  jet opening half-angle.
These energies are much smaller than the typical values of GRBs.
The other properties of GRB\,980425 are also unusual;
the large low-energy flux \citep{fro00a}, 
the low variability \citep{frr00},
the long spectral lag \citep{norris00}, 
and the slowly decaying X-ray afterglow \citep{pian00,pian03}.

Previous works suggest that the above peculiar observed properties 
may be explained
if the standard jet is seen from the off-axis viewing angle
(e.g.\citep{in01,na01}).
Following this scenario, the relativistic beaming effect 
reduces $E_{iso}$ and hence $E_\gamma$.
In this paper, in order to explain all of
the observed properties of GRB\,980425,
we reconsider the prompt emission of this event
using our simple jet model \citep{yin02,yin03a,yin03b,yyn03}.

\section{SPECTRAL ANALYSIS OF GRB\,980425 USING BATSE DATA}
\label{sec:analysis}
We argue the time-averaged observed spectral properties of 
GRB\,980425.
Using the BATSE data of GRB\,980425,
we analyze the spectrum within the time of FWHM
 of the peak flux in the light curve of 
BATSE channel\,2.
We fit the observed spectrum with the Band function.
The best-fit values are
$\alpha=-1.0\pm0.3$, $\beta=-2.1\pm0.1$, and
$E_p=54.6\pm20.9$\,keV, which
 are consistent with those derived by the previous works 
\citep{fro00a,ga98}.
This spectral property is similar to one of the
recently identified class of the X-ray flash (XRF)
\citep{He01a,ki02}.
The observed fluence of the entire emission is
$S$(20--2000\,keV) $=(4.0\pm0.74)\times10^{-6}$\,erg\,cm$^{-2}$,
thus we find
$E_{iso}=(6.4\pm1.2)\times10^{47}$\,ergs. 
The fluence ratio is 
$R_s=S$(20--50\,keV)/$S$(50--320\,keV) $=0.34\pm 0.036$.

\section{MODEL OF PROMPT EMISSION OF GRB\,980425}
We use a simple jet model of prompt emission of GRBs, where 
an instantaneous emission of infinitesimally thin shell is adopted
\citep{in01,yin02,yin03a,yin03b,yyn03}.
See \citet{yyn03} for details. 
We fix model parameters as $\alpha_B=-1$, $\beta_B=-2.1$,
$\gamma\nu'_0=2600\,{\rm keV}$, and $\gamma=100$.
Normalization of emitted luminosity is determined
so that $E_\gamma$ be observationally preferred value of
$1.15\times10^{51\pm0.35}(h/0.7)^{-2}$\,ergs \citep{bloom03}
when we see the jet from the on-axis viewing angle $\theta_v=0$.
Our calculations show that on-axis intrinsic peak energy becomes 
$E_p^{(\theta_v=0)}(1+z)\sim1.54\gamma\nu'_0\sim 4.0$~MeV,
in order to reproduce the observed quantities of GRB\,980425.
Indeed, there are some GRBs with higher intrinsic $E_p$; 
for example, $E_p(1+z)\sim2.0$\,MeV for GRB\,990123 
and 3.6\,MeV for GRB\,021004 \citep{amati02,bar03}.

The left panel of Figure.\,1 shows
$E_{iso}$ as a function of the viewing angle $\theta_v$.
When $\theta_v\lesssim\Delta\theta$,
$E_{iso}$ is constant, while for $\theta_v\gtrsim\Delta\theta$,
$E_{iso}$ is considerably smaller than the typical value of
$\sim10^{51-53}$\,ergs because of the relativistic beaming effect.

We next calculated $E_p$ and $R_s$
for the set of $\Delta\theta$ and $\theta_v^\ast$ that reproduces 
the observed $E_{iso}$ of GRB\,980425.
%
%
%
%
%
%
For our parameters, 
$\Delta\theta$ should be  between $\sim$18$\degr$ and 
$\sim$31$\degr$,
and then $\theta_v^\ast$ ranges between $\sim$24$\degr$ 
and $\sim$35$\degr$
in order to reproduce the observation results.
Thorough discussions on the right panel of Figure\,1 is found
in \citep{yyn03}.

\begin{figure}
\includegraphics[width=0.45\textwidth,height=.27\textheight]
                {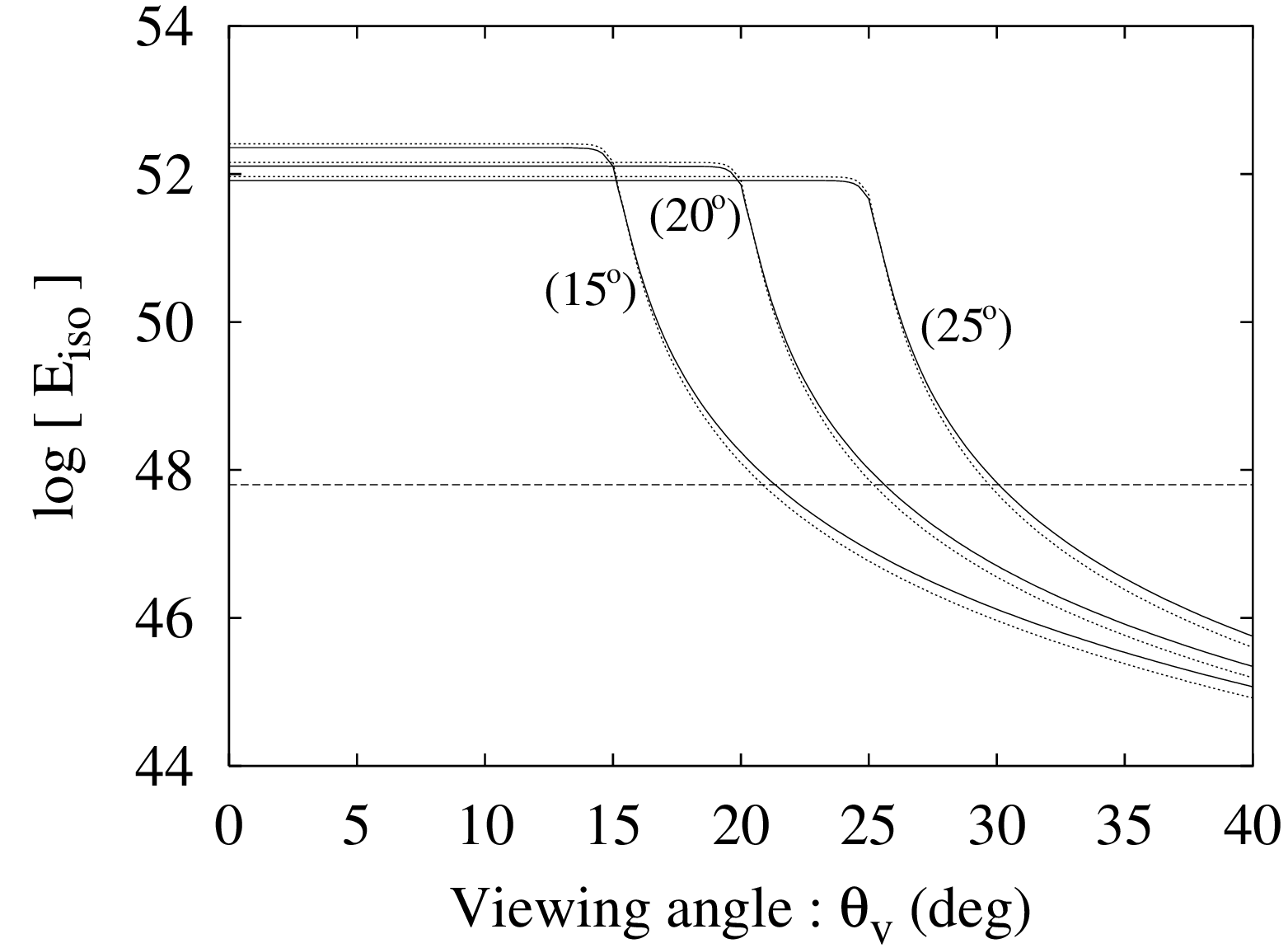}
\hspace{6mm}
\includegraphics[width=0.40\textwidth,height=.3\textheight]
                {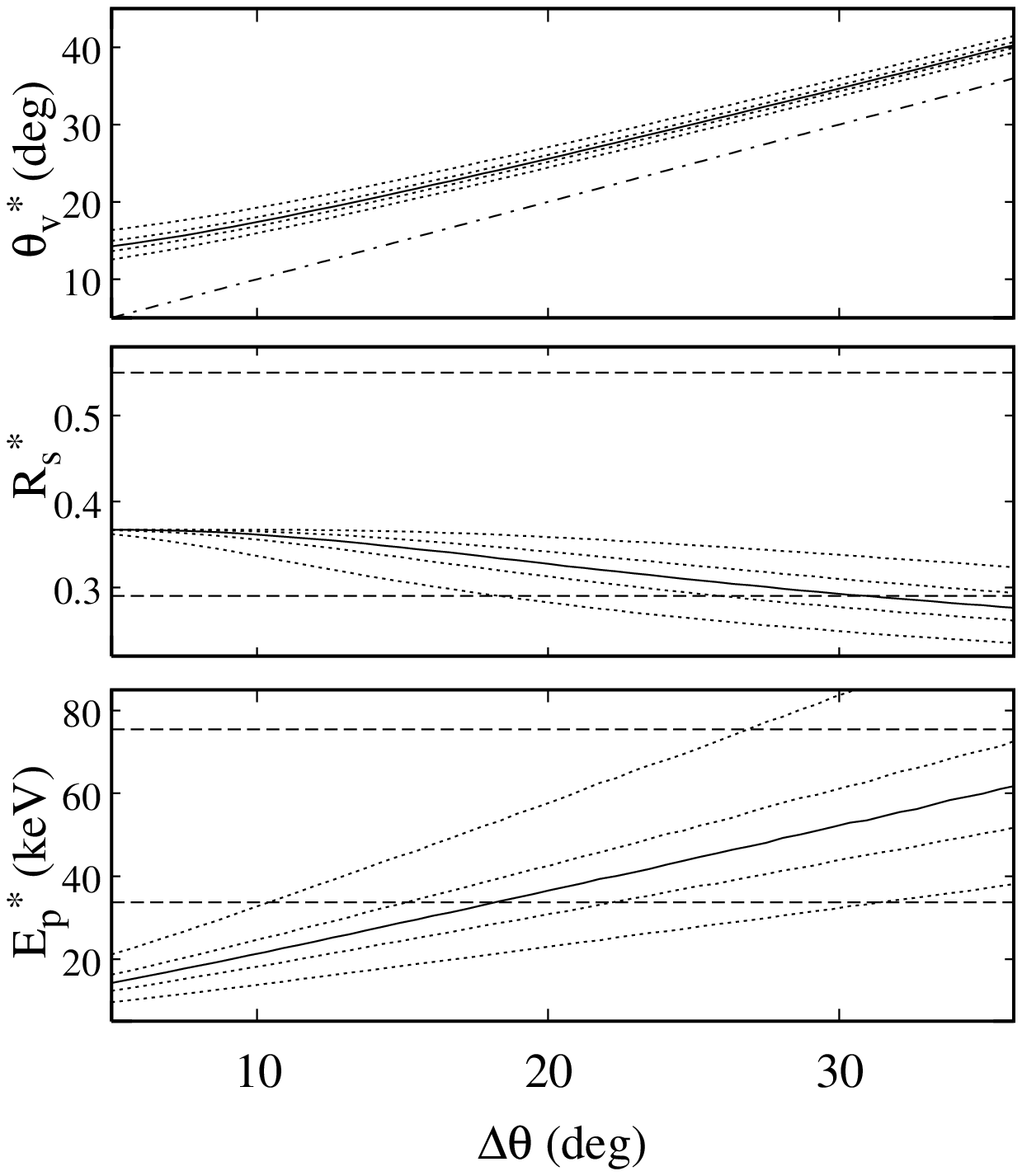}
\caption{
(Left panel):
the isotropic equivalent $\gamma$-ray energy $E_{iso}$
is shown as a function of the viewing angle $\theta_v$ for a
fixed jet opening half-angle $\Delta\theta$.
The source is located at $z=0.0085$.
The values of $\Delta\theta$ are shown in parentheses.
Solid lines correspond to the case of $\gamma\nu'_0=2600$~keV,
while dotted lines $\gamma\nu'_0=1300$~keV.
Horizontal dashed line represents the observed value of GRB\,980425.
(Right panel):
the upper panel shows $\theta_v^\ast$ for which
$E_{iso}$ is the  observed value of GRB\,980425,
while the middle and the lower panels represent the fluence ratio
$R_s^\ast=R_s^{(\theta_v=\theta_v^\ast)}$ and the peak energy 
$E_p^\ast=E_p^{(\theta_v=\theta_v^\ast)}$, respectively. 
Solid lines correspond to the fiducial case.
The dotted lines represent regions where $E_{iso}$ becomes
$(6.4\pm1.2)\times10^{47}$~ergs when
$\E_\gamma$ is in 1~$\sigma$ and 5~$\sigma$ level
around the fiducial value, respectively.
The dot-dashed line in the upper panel represents 
$\theta_v^\ast=\Delta\theta$.
Horizontal dashed lines in the middle and the lower panels represent 
the observational bounds.
}
\end{figure}

\section{DISCUSSION}\label{sec:dis}
We have found that when the jet of opening half-angle of 
$\Delta\theta\sim10$--30$\degr$ is seen from the 
off-axis viewing angle of $\theta_v\sim\Delta\theta+6\degr$,
observed quantities can be well explained.
%
%
Observed low variability can be explained
since only  subjets at the edge of the cone contribute to 
the observed quantities \citep{yin02}.
If the time unit parameter $r_0/c\beta \gamma^2$ is about 3\,sec,
which is in the reasonable parameter range,
the spectral-lag of GRB\,980425 can be also explained.

Our result might be able to explain the slowly decaying X-ray
 afterglow of GRB\,980425.
If we assume the density profile of ambient matter as
$n=n_0 (r/r_{ext})^{-2}$ with
$n_0r_{ext}^2=4\times10^{17}$\,cm$^{-2}$,
the break in the afterglow light curve should occur at
$t_b = 3.1\times10^2$\,days\,$E_{51}(\Theta/0.4\,{\rm rad})^2$, 
where $\Theta$ is defined by $\Theta^2=(\Delta\theta)^2+\theta_v^2$,
and $E$ is the total energy in the collimated jet \cite{na99,na01}.
Since our calculation suggests $\Theta$
should range between 0.4 and 0.67\,rad,
$t_b$ is consistent with the observation \citep{na01}.
Up to the break time, one can estimate 
the flux in the X-ray band as 
$F$(2--10\,keV)~$\propto t^{-0.2}$,
where we assume $\theta_v\gg\Delta\theta$ and the spectral index
of accelerated electrons as $p=2.2$ \citep{na99,na01}.
This result is also consistent with the observation
\citep{pian00,pian03}.
Furthermore,
the adopted value of $n_0r_{ext}^2$ corresponds to the mass loss
rate of the progenitor star 
$\dot{M}=1.3\times10^{-6}M_\odot$\,yr$^{-1}$\,$(v_{\rm W}/10^3\,
{\rm km}\,{\rm s}^{-1})$,
which might be able to explain the radio data (see \citep{wa03}).

The observed quantities of small $E_p$ and large fluence ratio
$R_s$  are the typical
values of the XRF \citep{fro00a,He01a,ki02}.
The operational definition of the {\it BeppoSAX}-XRF
is a fast X-ray transient with duration 
less than $\sim10^3$~s which is detected by WFCs 
and not detected by the GRBM.
If the distance to the source of GRB\,980425 were
larger than $\sim90$~Mpc, the observed flux in the $\gamma$-ray
band would have been less than the limiting sensitivity of GRBM,
so that the event would have been detected as an XRF. 

We might be able to explain the origin of a class with low 
$E_\gamma$ such as GRB\,980326 and GRB\,981226 \citep{bloom03}, 
and GRB\,030329 whose $E_\gamma$
is about $\sim 5 \times 10^{49}$ergs
if the jet break time of $\sim0.48$ days 
is assumed \citep{tamagawa,vanderspek}.
Let us consider the jet seen from a viewing angle
$\theta_v\sim\Delta\theta+\gamma_i^{-1}$,
where $\gamma_i$ is the Lorentz factor of a prompt $\gamma$-ray 
emitting shell.
Due to the relativistic beaming effect, observed $E_\gamma$ of such
a jet becomes an order of magnitude smaller than the standard energy.
At the same time, the observed peak energy $E_p$ is small
because of the relativistic Doppler effect.
In fact, the observed $E_p$ of the above three bursts are
less than $\sim$\,70\,keV.
In our model the fraction of low-$E_\gamma$ GRBs
becomes $2/(\gamma_i\Delta\theta)\sim 0.1$ since
the mean value of $\Delta \theta \sim 0.2$, 
while a few of them are observed
in $\sim$\,30 samples \citep{bloom03}.
In later phase, the Lorentz factor of afterglow emitting shock
$\gamma_f$ is smaller than $\gamma_i$, so that
$\theta_v<\Delta\theta+\gamma_f^{-1}$.
Then, the observed properties of afterglow may be similar to  the 
on-axis case $\theta_v\ll\Delta\theta$;
hence the observational estimation of the jet break time and the 
jet opening angle remains the same.

\begin{theacknowledgments}
This work was supported in part by
Grant-in-Aid for Scientific Research 
of the Japanese Ministry of Education, Culture, Sports, Science
and Technology, No.05008 (R.Y.), 
No.14047212 (T.N.), and No.14204024 (T.N.).
\end{theacknowledgments}


\IfFileExists{\jobname.bbl}{}
 {\typeout{}
  \typeout{******************************************}
  \typeout{** Please run "bibtex \jobname" to optain}
  \typeout{** the bibliography and then re-run LaTeX}
  \typeout{** twice to fix the references!}
  \typeout{******************************************}
  \typeout{}
 }


\begin{thebibliography}{}

\bibitem[Amati et al.(2002)]{amati02}
Amati,~L., et al. 2002, A\&A, 390, 81
%
%
%
\bibitem[Barraud et al.(2003)]{bar03}
Barraud,~C., et al. 2003, A\&A, 400, 1021
%
\bibitem[Bloom et al.(2003)]{bloom03}
Bloom,~J.S., et al. 2003, ApJ, 594, 674
%
%
%
%
%
%
\bibitem[Della~Valle et al.(2003)]{dv03}
Della~Valle,~M. et al. 2003, A\&A, 406, L33
%
%
%
%
\bibitem[Fenimore \& Ramirez-Ruiz(2000)]{frr00}
Fenimore,~E.~E. \& Ramirez-Ruiz.~E., 2000, astro-ph/0004176
%
%
\bibitem[Frontera et al.(2000a)]{fro00a}
Frontera,~F. et al. 2000a, ApJS, 127, 59
%
%
\bibitem[Galama et al.(1998)]{ga98}
Galama,~T.J., et al. 1998, Nature, 395, 670
%
%
%
%
%
%
%
\bibitem[Heise et al.(2001)]{He01a} 
Heise,~J., et al. 2001, in Proc. 2nd Rome Workshop:
GRBs in the Afterglow Era, astro-ph/0111246
%
%
%
\bibitem[Ioka \& Nakamura(2001)]{in01} 
Ioka,~K., \& Nakamura,~T. 2001, ApJ, 554, L163
%
%
%
%
%
%
\bibitem[Kippen et al.(2002)]{ki02} 
Kippen,~R.~M., et al. 2002, 
in Proc. Woods Hole Gamma-Ray Burst Workshop,
astro-ph/0203114
%
%
\bibitem[Kulkarni et al.(1998)]{kul98}
Kulkarni,~S.R., et al. 1998, Nature, 395, 663 
%
%
%
%
%
%
%
\bibitem[Nakamura(1999)]{na99}
Nakamura,~T. 1999, ApJ, 522, L101
%
%
\bibitem[Nakamura(2001)]{na01}
Nakamura,~T. 2001, Prog.~Theor.~Phys.~Suppl, 143, 50
%
\bibitem[Norris et al. (2000)]{norris00}
Norris,~J.P., Marani,~G.F., \& Bonnell,~J.T. 2000, ApJ, 534, 248
%
%
\bibitem[Pian et al.(2000)]{pian00}
Pian,~E. et al, 2000, ApJ, 536, 778
%
\bibitem[Pian et al.(2003)]{pian03}
Pian,~E. et al, 2003, astro-ph/0304521
%
%
%
%
%
%
%
%
%
%
%
%
%
\bibitem[Stanek et al.(2003)]{stanek}
Stanek, K.Z. et al. 2003, ApJ, 591, L71
%
%
%
\bibitem[Tamagawa et al.(2003)]{tamagawa}
Tamagawa,~T. et al. 2003, in this proceeding
%
\bibitem[Vanderspek(2003)]{vanderspek}
Vanderspek,~R. et al. 2003, GCN circ. 1997
%
%
\bibitem[Waxman(2003)]{wa03}
Waxman,~E.\ 2003, astro-ph/0310320
%
%
%
%
\bibitem[Yamazaki, Ioka \& Nakamura(2002)]{yin02} 
Yamazaki,~R., Ioka,~K., \& Nakamura, T. 2002, ApJ, 571, L31
%
\bibitem[Yamazaki, Ioka \& Nakamura(2003a)]{yin03a} 
Yamazaki,~R., Ioka,~K., \& Nakamura, T. 2003a, ApJ, 591, 283
%
\bibitem[Yamazaki, Ioka \& Nakamura(2003b)]{yin03b} 
Yamazaki,~R., Ioka,~K., \& Nakamura, T. 2003b, ApJ, 593, 941
%
\bibitem[Yamazaki et al.(2003)]{yyn03} 
Yamazaki,~R., Yonetoku,~D., \& Nakamura, T. 2003, ApJ, 594, L79

%
%

\end{thebibliography}
\end{document}